\documentclass[aps,twocolumn]{revtex4-1}
\usepackage{graphicx}
\usepackage{bm}
\usepackage{amsfonts}
\usepackage[usenames, dvipsnames]{color}
\usepackage[breaklinks]{hyperref}
\usepackage{epstopdf, epsfig}
\usepackage[toc,page]{appendix}

\graphicspath{{./Figs/}}

\begin{document}

\title{Machine learning strategies for path-planning
    microswimmers in turbulent flows}

\author{Jaya Kumar Alageshan$^\dagger$, Akhilesh Kumar Verma$^\dagger$, 
         J\'er\'emie Bec$^\ddagger$, and Rahul Pandit$^\dagger$}
\altaffiliation{also at Jawaharlal Nehru Centre for Advanced Scientific Research,
            Bangalore, India - 560064.}
\email{rahul@iisc.ac.in}
\email{$^\dagger$jayaka@iisc.ac.in}
\email{akhilesh@iisc.ac.in}
\email{$^\ddagger$jeremie.bec@mines-paristech.fr}
\affiliation{$^\dagger$Centre for Condensed Matter Physics, Department of Physics,
         Indian Institute of Science, Bangalore, 
             India - 560012.}
\affiliation{$^\ddagger$MINES ParisTech, PSL Research University, CNRS, CEMEF, 
       Sophia–Antipolis, France.} 

\begin{abstract}
We develop an \textit{adversarial-reinforcement} learning scheme for
microswimmers in statistically homogeneous and isotropic turbulent
fluid flows, in both two (2D) and three dimensions (3D). We show that
this scheme allows microswimmers to find non-trivial paths, which enable
them to reach a target on average in less time than a na\"ive
microswimmer, which tries, at any instant of time and at a given position in 
space, to swim in the direction of the target. We 
use pseudospectral direct numerical simulations (DNSs) of the 2D and 3D 
(incompressible)
Navier-Stokes equations to obtain the turbulent flows.  We then
introduce passive microswimmers that try to swim along a given
direction in these flows; the microswimmers do not affect the flow,
but they are advected by it. Two, non-dimensional, control parameters
play important roles in our learning scheme: (a) the ratio
$\tilde{V}_s$ of the microswimmer's bare velocity $V_s$ and the
root-mean-square (rms) velocity $u_{rms}$ of the turbulent fluid; and (b) the
product $\tilde{B}$ of the microswimmer-response time $B$ and the rms 
vorticity $\omega_{rms}$ of the fluid. We show that
the average time required for the microswimmers
to reach the target, by using our adversarial-learning scheme, eventually
reduces below the average time taken by microswimmers that follow the
na\"ive strategy. 
\end{abstract}

\maketitle

\section{Introduction}
Machine-learning techniques and advances in computational facilities have 
led to significant improvements in obtaining solutions to optimization problems,
e.g., to problems in path planning and optimal transport, referred to in control systems
as Zermelo's navigation problem~\cite{Zermelo}. With vast amounts of data available
from experiments and simulations in fluid dynamics, machine-learning techniques
are being used to extract information that is useful to control and optimize 
flows~\cite{ML_FM}.
Recent studies include the use of reinforcement learning, in fluid-flow
settings, e.g., (a) to optimise the soaring of a glider in thermal
currents~\cite{Soaring} and (b) the development of an optimal scheme
in two- (2D) and three-dimensional (3D) fluid flows that are time
independent~\cite{PRL,3D}. 
Optimal locomotion, in response to stimuli, is also important in biological
systems ranging from cells and micro-organisms~\cite{Dusenbery,Durham,Michalec} to birds, animals, and fish~\cite{Fish}; such locomotion is often termed \textit{ taxis}~\cite{Barrows}.

It behooves us, therefore, to explore machine-learning strategies for optimal
path planning by microswimmers in turbulent fluid flows. We initiate such a
study for microswimmers in 2D and 3D turbulent flows. In particular, we
consider a dynamic-path-planning problem that seeks to minimize the average time taken
by microswimmers to reach a given target, while moving in a turbulent fluid
flow that is statistically homogeneous and isotropic. We
develop a novel, multi-swimmer, \textit{adversarial-$\mathcal{Q}$-learning} algorithm to
optimise the motion of such microswimmers that try to swim towards a specified
target (or targets). Our adversarial-$\mathcal{Q}$-learning approach ensures that the
microswimmers perform at least as well as those that adopt the following
na\"ive strategy: at any instant of time and at a given position in space, a
na\"ive microswimmer tries to point in the direction of the target.  We examine
the efficacy of this approach as a function of the following two dimensionless
control parameters: (a) $\tilde{V}_s = V_s/u_{rms}$, where the microswimmer's
bare velocity is $V_s$ and the the turbulent fluid has the root-mean-square
velocity $u_{rms}$; and (b) $\tilde{B}=B\;\omega_{rms}$, where $B$ is the
microswimmer-response time and $\omega_{rms}$ the rms vorticity of the 
fluid. We show, by extensive
direct numerical simulations (DNSs), that the average time $\langle T \rangle$, 
required by a microswimmer to reach a target at a fixed distance, is lower, if it 
uses our adversarial-$\mathcal{Q}$-learning scheme, than if it uses the na\"ive 
strategy.

\section{Background flow and microswimmer dynamics}

For the low-Mach-number flows we consider, 
the fluid-flow velocity $\mathbf{u}$ satisfies  the incompressible
Navier-Stokes (NS) equation. In two dimensions (2D), we write the NS
equations in the conventional vorticity-stream-function form, which accounts
for incompressibility in 2D~\cite{RP_Review}: 
 \begin{equation}
 \left(\partial_t + {\mathbf u}\cdot \nabla\right) \omega = \nu \nabla^2 \omega - \alpha \;
          \omega+ F_\omega ;
 \label{2DNS} 
 \end{equation}
here, $\mathbf{u}\equiv(u_x,u_y)$ is the fluid velocity, $\nu$ is the kinematic viscosity,
$\alpha$ is the coefficient of friction (present in 2D, e.g., because of air drag
or bottom friction) and the vorticity $\omega = (\nabla \times \mathbf{u})$, which is
normal to $\mathbf{u}$ in 2D. The 3D incompressible NS equations are
\begin{eqnarray}
 \left(\partial_t + {\mathbf{u}}\cdot \nabla\right) \mathbf{u} 
     &=& -\nabla p/\rho+\mathbf{f}+ \nu \nabla^2 {\bf u} ;  \nonumber\\
	\nabla . {\bf u} &=& 0;
 \label{3DNS} 
 \end{eqnarray}
$p$ is the pressure and the density $\rho$ of the incompressible fluid is taken 
to be $1$; 
the large-scale forcing $F_\omega$ (large-scale random forcing in 2D)
or  $\mathbf{f}$ 
(constant energy injection in 3D) maintains the statistically steady, homogeneous, and 
isotropic turbulence, for which it is natural to use periodic boundary conditions.

%

We consider a collection of $\mathcal{N}_p$ passive, non-interacting
microswimmers in the turbulent flow; $\mathbf{X}_i$ and $\hat{\mathbf{p}}_i$ are the position and swimming direction of the microswimmer.
Each microswimmer is assigned a target located at $\mathbf{X}^{T}_i$. We are 
interested in 
minimizing the time $\mathcal{T}$ required by a microswimmer, which is released at a distance 
$r_0 = |\mathbf{X}_i(0)-\mathbf{X}^T_i|$ from its target, to approach within a small distance 
$r = |\mathbf{X}_i(\mathcal{T})-\mathbf{X}^T_i| \ll r_0$ of this target. The microswimmer's
position and swimming direction evolve as follows~\cite{Pedley}:
\begin{eqnarray}
	\frac{d\mathbf{X}_i}{dt} &=& \mathbf{u}(\mathbf{X}_i,t) + V_s \; \hat{\mathbf{p}}_i \; ; \label{eq:Swimmer_Evolution1}\\
	\frac{d\hat{\mathbf{p}}_i}{dt} &=& \frac{1}{2B} \left[\hat{\mathbf{o}}_i-(\hat{\mathbf{o}}_i
	     . \hat{\mathbf{p}}_i) \; \hat{\mathbf{p}}_i \right] + \frac{1}{2} {\mathbf{\omega}} \times
	      \hat{\mathbf{p}}_i \; ;
	\label{eq:Swimmer_Evolution2}
\end{eqnarray}
here, we use bi-linear (tri-linear) interpolation in 2D (3D) to determine the
fluid velocity $\mathbf{u}$ at the microswimmer's position $\mathbf{X}_i$ from
eq.~\ref{3DNS}; $V_s \hat{\mathbf{p}}_i$ is the swimming velocity, $B$ is the
time-scale associated with the microswimmer to align with the flow, and $\hat{\mathbf{o}}_i$ is the control direction.
Equation~\ref{eq:Swimmer_Evolution2} implies that $\hat{\mathbf{p}}_i$ tries
to align along $\hat{\mathbf{o}}_i$.  We define the following non-dimensional
control parameters: $\tilde{V}_s = V_s/u_{rms}$, where $u_{rms} = \langle
|\mathbf{u}|^2 \rangle^{1/2}$ is the root-mean-square ($rms$) fluid flow
velocity, and $\tilde{B} = B/\tau_{\Omega}$, where 
$\tau_\Omega = \omega_{rms}^{-1}$;
$\omega_{rms}=\langle|\omega|^2\rangle^{1/2}$ denotes the root-mean-square vorticity.

\section{Adversarial $\mathcal{Q}$-learning for smart microswimmers}

\begin{figure}[!ht]
	\includegraphics[scale=0.7]{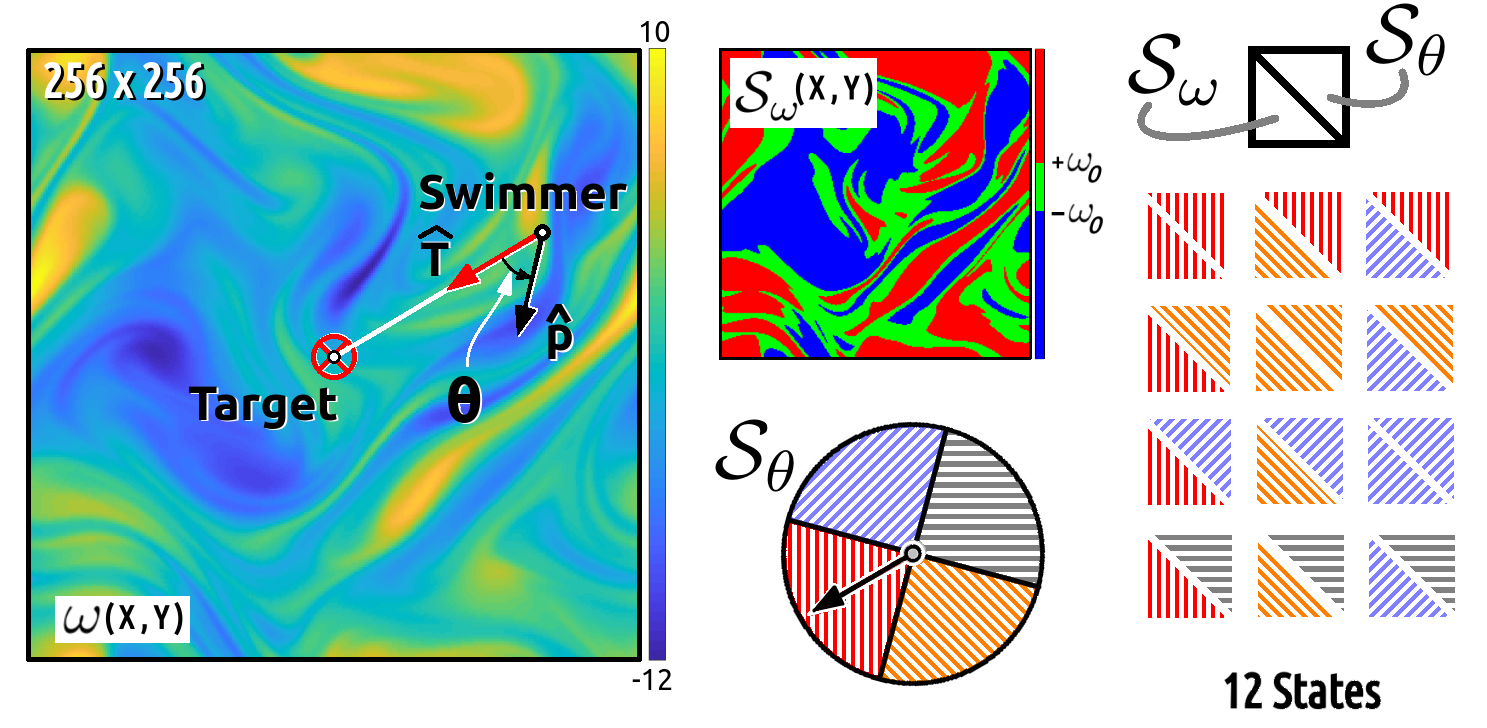}
	\caption{Left panel: a pseudocolor plot of the vorticity
	         field, with a microswimmer represented by a small white circle; the
	         black arrow on the microswimmer indicates its swimming direction, 
	         $\hat{\mathbf{p}}$, the red arrow represents the direction towards the 
	         target, $\hat{\mathbf{T}}$, and $\theta$ is the angle between 
	         $\hat{\mathbf{p}}$ and $\hat{\mathbf{T}}$.
	         Top-center panel shows the discretized vorticity
	         states (red $|||$: $\omega>\omega_0$, green 
	         ${\backslash}\backslash\backslash$: 
	         $-\omega_0\leq\omega\leq\omega_0$,
	         blue $///$: $\omega<-\omega_0$). In our approach we use $\omega_0 
	         = \omega_{rms}$. The bottom-center panel indicates the color
	         code for the discretized $\theta$(red $|||$: $-\pi/4\leq \theta <\pi/4$;
	         orange ${\backslash}\backslash\backslash$: $\pi/4\leq \theta < 3\pi/4$; 
	         blue $///$: $-3\pi/4 \leq \theta < -\pi/4$;
	         gray $\equiv$: $3\pi/4 \leq \theta < 5\pi/4$). The right panel lists all 
	         possible discrete states of the microswimmers, via colored squares
	         where the lower half stands for the vorticity state, $\mathcal{S}_\omega$,
	         and the upper half represents the direction state, $\mathcal{S}_\theta$
	         .}
	\label{fig:State_Color}
\end{figure}

Designing a strategy consists in choosing appropriately the control direction 
$\hat{\mathbf{o}}_i$, as a function of the instantaneous state of the microswimmer,
in order to minimize the mean arrival time $\langle \mathcal{T} \rangle$. To develop a
\textit{tractable} framework for $\mathcal{Q}$-learning, we use a \textit{finite 
number of states} by discretizing the fluid vorticity $\omega$ at the microswimmer's 
location into 3 ranges of values labelled by $\mathcal{S}_\omega$ and the angle 
$\theta_i$, between $\hat{\mathbf{p}}_i$ and $\hat{\mathbf{T}}_i$, into 4 ranges 
$\mathcal{S}_\theta$, as shown in fig~\ref{fig:State_Color}. The choice of  
$\hat{\mathbf{o}}_i$ is then reduced to a map from $\left(\mathcal{S}_\omega, 
\mathcal{S}_\theta \right)$ to an \textit{action} set, $\mathcal{A}$, which we 
also discretize into the following four possible actions: $\mathcal{A} := 
\left\{ \hat{\mathbf{T}}_i, -\hat{\mathbf{T}}_i, \hat{\mathbf{T}}_{i\perp}, 
-\hat{\mathbf{T}}_{i\perp} \right\}$, where $\hat{\mathbf{T}}_i = 
(\mathbf{X}^T_i-\mathbf{X}_i)/|\mathbf{X}^T_i-\mathbf{X}_i|$ is the unit vector
pointing from the swimmer 
to its target and $(\hat{\mathbf{T}}_{i\perp} \cdot \hat{\mathbf{T}}_i) = 0$.  
Therefore, for the na\"ive strategy $\hat{\mathbf{o}}_i(s_i) \equiv 
\hat{\mathbf{T}}_i$, $\forall \; s_i \in (\mathcal{S}_\omega, \mathcal{S}_\theta)$. 
This strategy is optimal if $\tilde{V}_s \gg 1$: Microswimmers have an almost ballistic 
dynamics and move swiftly to the target. For $\tilde{V}_s \simeq 1$, vortices affect 
the microswimmers substantially, so we have to develop a nontrivial 
$\mathcal{Q}$-learning strategy, in which $\hat{\mathbf{o}}_i$ is a 
function of ${\mathbf{\omega}} ({\mathbf{X}_i},t)$ and $\theta_i$.

In our $\mathcal{Q}$-learning scheme, we assign a {\it quality} value to each
state-action binary relation of microswimmer $i$ as follows:
$\mathcal{Q}_i:(s_i,a_i)\rightarrow \mathbb{R}$, where $s_i \in
(\mathcal{S}_\omega, \mathcal{S}_\theta)$ and $a_i \in \mathcal{A}$;  
and we use the $\epsilon$-greedy method~\cite{Watkins} (with parameter $\epsilon_g$), 
in which the control direction is chosen from
the probability distribution $\mathcal{P}\left[\hat{\mathbf{o}}_i(s_i)\right]
= \epsilon_g/4 + (1-\epsilon_g) \; \delta\left(\hat{\mathbf{o}}_i(s_i)-
\hat{\mathbf{o}}_{max}\right)$, where $\hat{\mathbf{o}}_{max} := \mathop{\mathrm{argmax}}_{a
\in \mathcal{A}} \mathcal{Q}_i(s_i,a)$ and $\delta(.)$ is the Dirac delta function.
At each iteration, $\hat{\mathbf{o}}_i$ is calculated as above and
the microswimmer evolution is performed by using
eqs.~\ref{eq:Swimmer_Evolution1} and~\ref{eq:Swimmer_Evolution2}. In the canonical
$\mathcal{Q}$-learning approach, during the learning process, each of the
$\mathcal{Q}_i$'s are evolved by using the Bellman equation~\cite{Sutton}
below, whenever there is a state change, {\it i.e.,} $s_i(t) \neq s_i(t+\delta
t)$:

\begin{eqnarray}
\mathcal{Q}_i\left( s_i(t), \hat{\mathbf{o}}_i(s_i(t)) \right) \; \mapsto \; (1-\lambda)\; 
          \mathcal{Q}_i\left( s_i(t), \hat{\mathbf{o}}_i(s_i(t)) \right) 
           \nonumber  \\ 
           \;\;\;\; + \;
  \lambda\,\left[ \mathcal{R}_i(t)+ \gamma\,\max_{a \in \mathcal{A}} 
               \mathcal{Q}_i(s_i(t+\delta t), a ) \right], \label{eqn:Q}
\end{eqnarray}
where $\lambda$ and $\gamma$ are learning parameters that are set to optimal
values after some numerical exploration (see tab.~\ref{tab:Parameters}), and
$\mathcal{R}_i$ is the reward function. For the path-planning problem we define
$\mathcal{R}_i(t) = | \mathbf{X}_i(t-n \: \delta t) - \mathbf{X}^T_i| - |
\mathbf{X}_i(t) - \mathbf{X}^T_i|$, where $n=\min_{l \in \mathbb{N}} \left\{
	s_i(t-l \; \delta t) \neq s_i(t) \right\}$.  According to
eq.~\ref{eqn:Q}, any $\hat{\mathbf{o}}_i$ for which $\mathcal{R}_i$ is
positive can be a solution, and there exist many such solutions that are
sub-optimal compared to the na\"ive strategy. 

\begin{table}[!ht]
    \begin{center}
	\begin{tabular}{c c c}
	    \hline
	    \hline
		$\gamma = 0.99 \;\;$ &  $\;\;$
		$\lambda = 0.01$   \\
		$\epsilon_g = 0.001 \;\;$ & $\;\;$
		$\omega_0 / \omega_{rms} = 1.0$ \\
		\hline
		\\
	\end{tabular}
	\caption{List of learning parameter values: $\gamma$ is the earning discount,
	          $\lambda$ is the learning rate, $\epsilon_g$ is the 
	          $\epsilon$-greedy algorithm parameter that represents the probability
	          with which the non-optimal action is chosen,
	          $\omega_0$ is the cut-off used for defining
	          $S_\omega$, and $\omega_{rms}$ is the rms value of $\omega$. }
	\label{tab:Parameters}
	\end{center}
\end{table}

\begin{figure}[!ht]
    \begin{center}
	\includegraphics[scale=0.25]{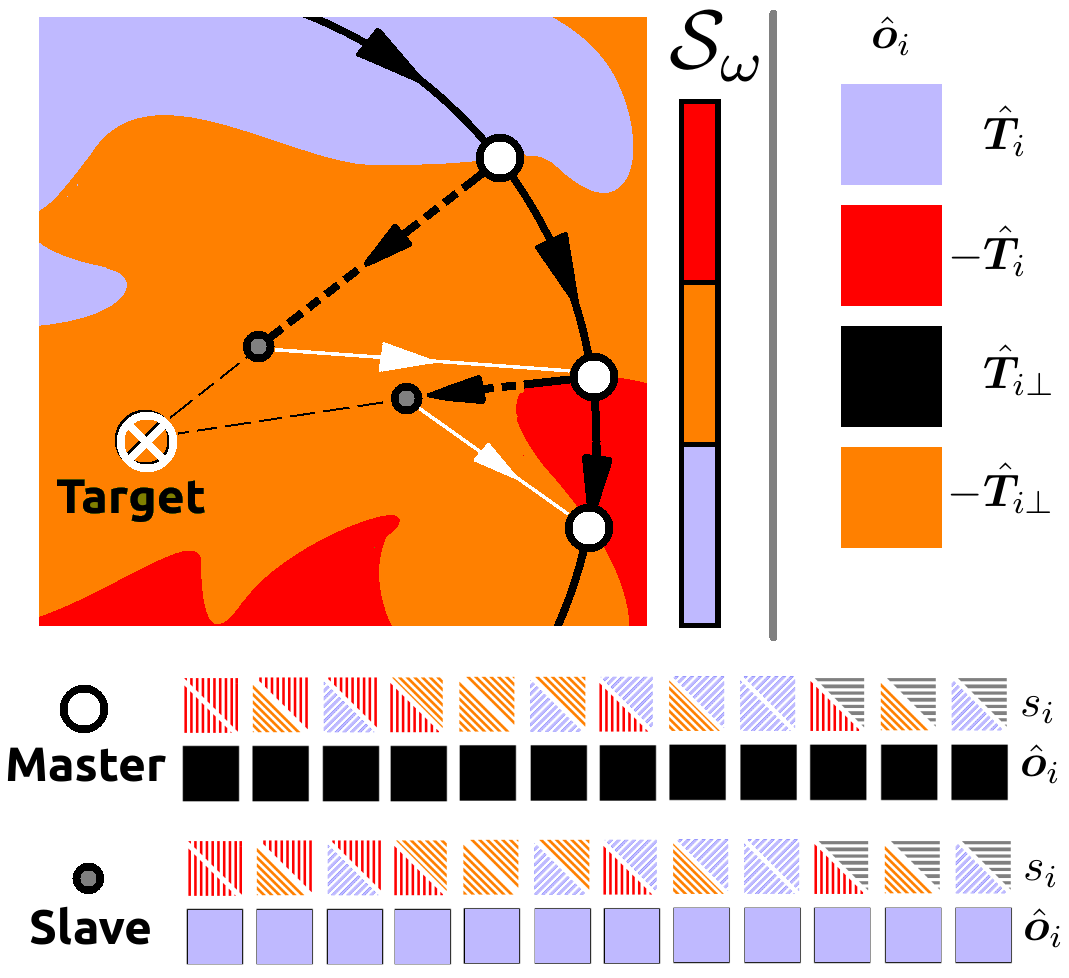}
	\caption{Top-left panel: a schematic diagram illustrating the trajectories of
	 master (black line) and slave (dashed black line) microswimmers superimposed 
	 on a pseudocolor plot of the two-dimensional (2D) discrete vorticity field 
	 $\mathcal{S}_\omega$; the master undergoes a state change at the points 
	 shown by white filled circles; white arrows indicate the re-setting of the
	 slave's trajectory. Top-right panel: color code for the control direction 
	 $\hat{o}_i$; for the states $s_i \in (\mathcal{S}_\omega,\mathcal{S}_\theta)$ 
	 see Fig.~\ref{fig:State_Color}. Bottom panel: control maps for the master and 
	 slave; for the 
	 purpose of illustration, we use $\hat{\mathbf{o}}_i = \hat{\mathbf{T}}_{i\perp}$, 
	 for the master; for $\tilde{V}_s \gg 1$ and $\tilde{B} = 0$, this leads 
	 to the circular path shown in our schematic diagram.}
	\label{fig:Master_Slave}
	\end{center}
\end{figure}

To reduce the solution space, we propose an {\it adversarial} scheme: Each
microswimmer, the {\it master}, is accompanied by a {\it slave} microswimmer, with
position $\mathbf{X}^{Sl}_i(t)$, that shares the same target at $\mathbf{X}^T_i$, and follows the na\"ive strategy, {\it i.e.}, $\hat{\mathbf{o}}^{Sl}_i(t) \equiv 
\hat{\mathbf{T}}^{Sl}_i = (\mathbf{X}^{Sl}_i - \mathbf{X}^T_i)/|\mathbf{X}^{Sl}_i -
\mathbf{X}^T_i|$.
Now, whenever the master undergoes a state change, the corresponding slave's
position and direction are re-initialized to that of the master, {\it i.e.}, if
$s_i(t) \neq s_i(t+\delta t)$, then $\mathbf{X}^{Sl}_i(t+\delta t) = \mathbf{X}_i(t+\delta t)$ and $\hat{\mathbf{p}}^{Sl}_i(t+\delta t) = \hat{\mathbf{p}}_i(t+\delta t)$ (see fig.~\ref{fig:Master_Slave}). Then the reward function
for the {\it master} microswimmer is given by $\mathcal{R}_i^{AD}(t) = |\mathbf{X}^{Sl}_i(t) -\mathbf{X}^T_i|-|\mathbf{X}_i(t)-\mathbf{X}^T_i|$; {\it i.e.},
only those changes that improve on the na\"ive startegy are favored. 

In the conventional $\mathcal{Q}$-learning approach~\cite{Watkins,Survey}, the
matrices $\mathcal{Q}_i$ of each microswimmer evolve independently; this  matrix is
updated only after a state change, so a large number of iterations are required
for the convergence of $\mathcal{Q}_i$. To speed-up this learning process, we
use the following multi-swimmer, parallel-learning scheme: all the microswimmers
share a common $\mathcal{Q}$ matrix, \textit{ i.e.}, $\mathcal{Q}_i =
\mathcal{Q}, \forall i$. At each iteration, we choose one microswimmer at random,
from the set of microswimmers that have undergone a state change, to update the
corresponding element of the $\mathcal{Q}$ matrix (flow chart in 
Appendix~\ref{sec:Flowchart}); this ensures that the $\mathcal{Q}$ matrix
is updated at almost every iteration and so it converges rapidly.

\section{Numerical simulation}

We use a pseudospectral DNS~\cite{canuto,pramanareview}, with the $2/3$
dealiasing rule to solve eqs.~\ref{2DNS} and~\ref{3DNS}. For time marching we use a 
third-order
Runge-Kutta scheme in 2D and the exponential Adams-Bashforth time-integration
scheme in 3D; the time step $\delta t$ is chosen such that the
Courant-Friedrichs-Lewy (CFL) condition is satisfied.
Table~\ref{tab:Flow_Parameters} gives the parameters for our DNSs in 2D and
3D, such as the number $N$ of collocation points and the Taylor-microscale
Reynolds numbers $R_\lambda = u_{rms} \lambda/\nu$, where the Taylor microscale
$\lambda = \left[ \sum_k k^2 E(k) / \sum_k E(k) \right]^{-1/2}$.

\begin{table}[!ht]
    \begin{center}
    	\caption{Parameters: $N$, the number of collocation points;
	          $\nu$ the kinematic viscosity; $\alpha$ the coefficient of 
		  friction; $\delta t$ the time step; and $R_\lambda$ the Taylor-microscale
          Reynolds number.}
	\begin{tabular}{c | c c}
	     & 2D & 3D \\
	    \hline
	    $N$ & $\;\;256 \times 256\;\;$ & $128 \times 128 \times 128$ \\
	    $\nu$ & $0.002$ &  $0.002$\\
		$\alpha$ & $0.05$ & $0.00$ \\
		$\delta t$ & $5\times 10^{-4}$ &  $8\times 10^{-3}$\\
		$R_{\lambda}$ & $130$ & $30$ \\
		\hline
	\end{tabular}
	\label{tab:Flow_Parameters}
	\end{center}
\end{table}

\subsection{Na\"ive microswimmers}

\begin{figure}[!ht]
  \begin{center}
    \includegraphics[width=\columnwidth]{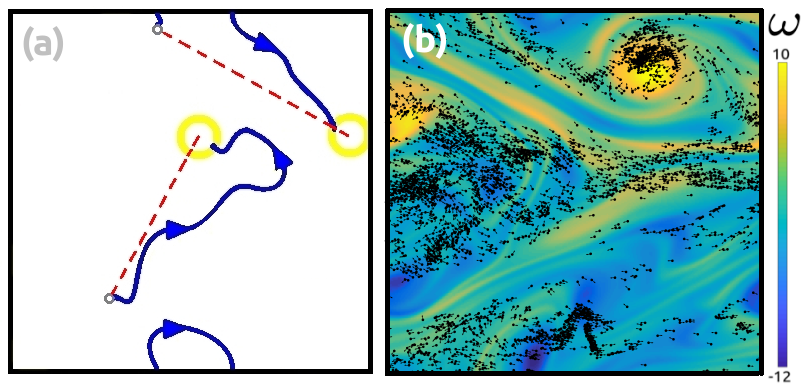}
  \end{center}
  \vspace{-18pt}
  \caption{\label{fig:Trap}
    (a) Illustrative (blue) paths for two microswimmers, with their corresponding
     (yellow) circular target regions (mapping in red dashed lines) where
     the microswimmer is eventually absorbed and re-initialized. We consider random 
     positions
     of targets and initialize a microswimmer at a fixed distance from its corresponding
     target with randomized $\hat{\mathbf{p}}$; (b) a snapshot of the microswimmer
     distribution, in a vorticity field ($\mathbf{\omega}$), for the na\"ive 
     strategy, at time $t = 30 \tau_{\Omega}$, with $\tilde{V}_s = 1$. Here,
     the initial distance of the microswimmers from their respective targets is $L/3$ 
	and the 
     target radius is $L/50$; we use a system size $L$ with periodic boundary 
	conditions in all directions.}
\end{figure}

\begin{figure}[!ht]
    \begin{center}
	\includegraphics[scale=0.09]{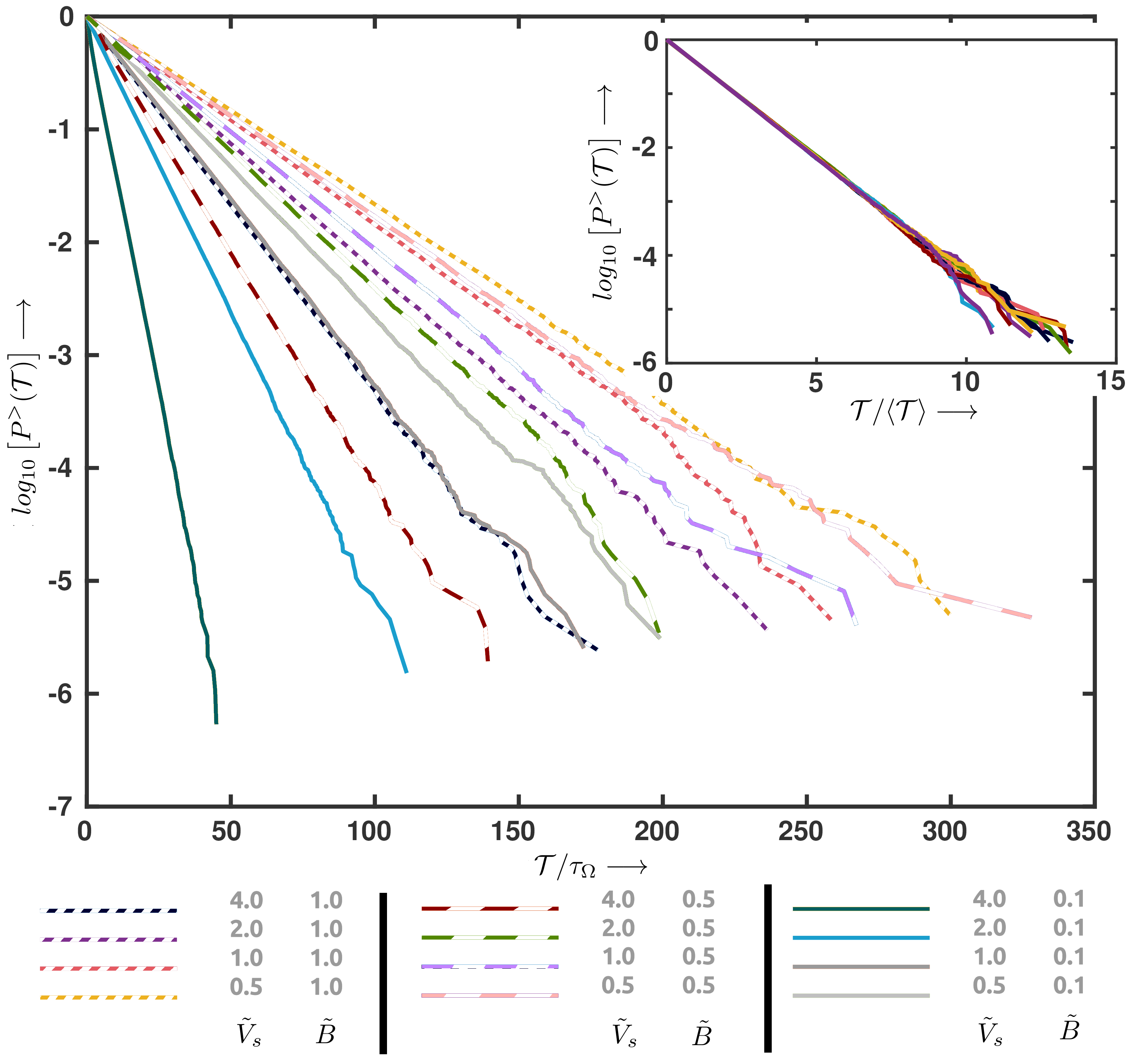}
	\caption{Plots showing exponential tails in $P^>(\mathcal{T})$ for the na\"ive strategy,
	         with different values of $\tilde{V}_s$ and $\tilde{B}$. 
	         The inset shows how these data collapse when, $\mathcal{T}$  is 
	         normalized, for each curve, by the corresponding $\langle \mathcal{T} \rangle$,
	         which implies $P^>(\mathcal{T}) \sim \exp\left(-\mathcal{T}/\langle 
	         \mathcal{T} \rangle\right)$.}
	\label{fig:Tail}
	\end{center}
\end{figure}

The average time taken by the microswimmers
to reach their targets is $\langle \mathcal{T} \rangle$ (see fig.~\ref{fig:Trap}).  If
$\hat{\mathbf{T}}_i = (\mathbf{X}_i - \mathbf{X}^T_i)/|\mathbf{X}_i -
\mathbf{X}^T_i|$ is the unit vector pointing from the microswimmer to the
target, then for $\tilde{V}_s\gg 1$ we expect the na\"ive strategy, \textit{i.e.},
$\hat{\mathbf{o}}_i = \hat{\mathbf{T}}_i$, to be the optimal one. For
$\tilde{V}_s\simeq 1$, we observe that the na\"ive strategy leads to the
trapping of microswimmers (fig.~\ref{fig:Trap}(b)) and gives rise to
exponential tails in the arrival-time ($\mathcal{T}$) probability distribution function (PDF);
in fig.~\ref{fig:Tail} we plot the associated complementary cumulative
distribution function (CCDF) $P^{>}(\mathcal{T}) = \int_T^{\infty} 
\wp(\tau) \; d\tau$, where $\wp(\tau) \: d\tau$ is the probability of particle arrival 
in the time interval $[\tau, \tau+d\tau]$ and $\tau$ is the time since initialization
of the microswimmer. As a consequence of trapping, $\langle \mathcal{T} \rangle$ is 
dominated by the exponential tail of the distribution, as can be seen from 
fig.~\ref{fig:Tail}.

\subsection{Smart microswimmers}

\begin{figure}[!ht]
  \centering
  \includegraphics[scale=0.4]{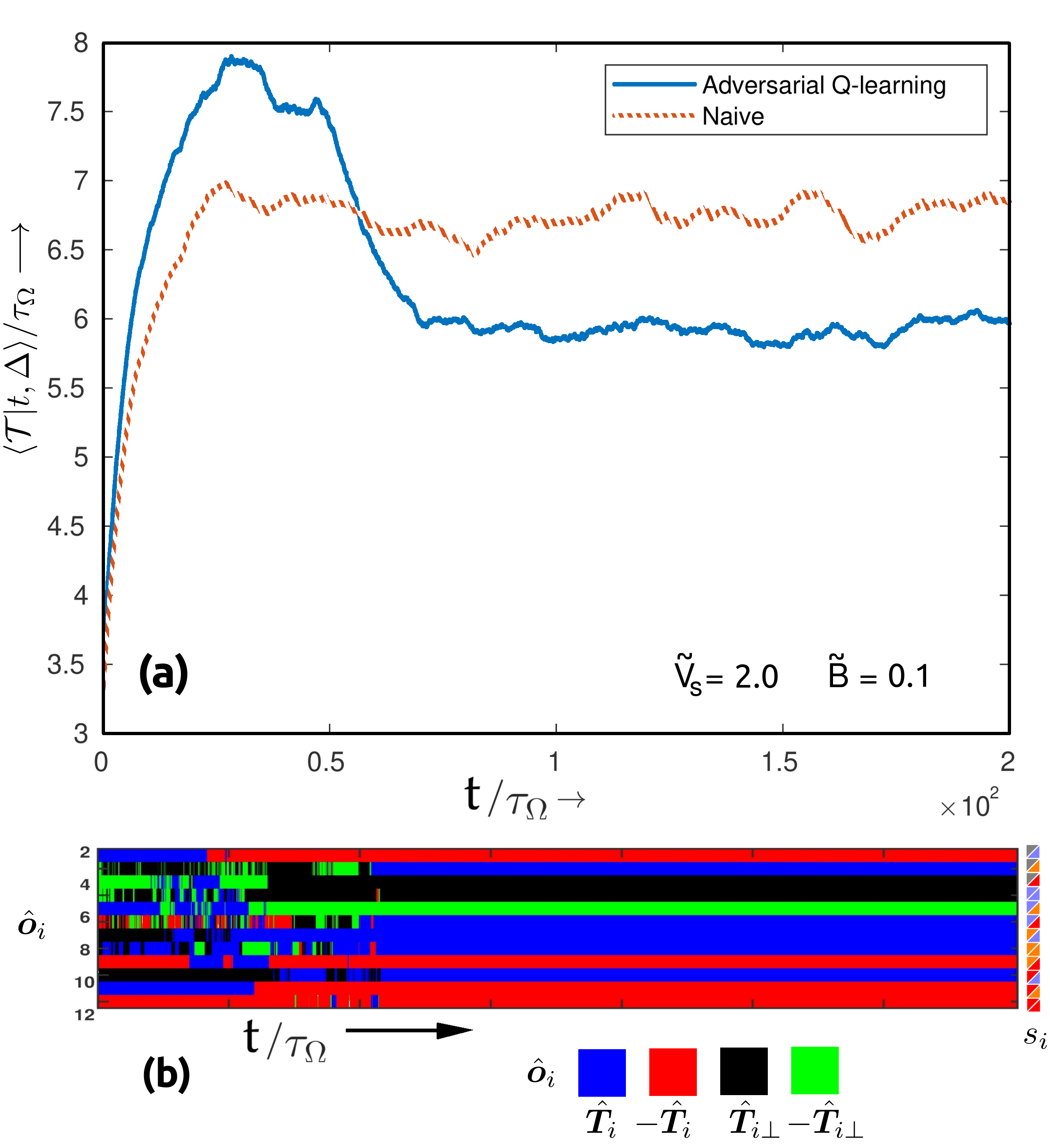}
  \caption{ Learning statistics: (a) Plot of $\langle \mathcal{T} | t, \Delta
             \rangle$, with $\Delta = 10\; \tau_{\Omega}$, in 2D. Adversarial 
             $\mathcal{Q}$-learning initially shows a transient behavior, before settling
             to a lower value of $\langle \mathcal{T} \rangle$ than that in the na\"ive
             strategy. (b) The evolution of the control map, $\hat{\mathbf{o}}_i$, where 
	         the color codes represent the actions that
	         are performed for each of the 12 states.  
	         Initially, $\mathcal{Q}$-learning explores different strategies and
	         settles down to a $\hat{\mathbf{o}}_i$ that shows, consistently, improved
	         performance relative to the na\"ive strategy. 
          }
   \label{fig:Performance}
\end{figure}

In our approach, the random initial positions of the microswimmers
ensures that they explore different states without reinitialization for each
epoch. Hence, we present results with 10000 microswimmers, for a single epoch. 
In our single-epoch approach, the control map
$\hat{\mathbf{o}}_i$ reaches a steady state once the learning process is
complete (fig.  \ref{fig:Performance}(b)). We would like to
clarify here that, in our study, the training is performed in the fully
turbulent time-dependent flow; even though this is more difficult than
training in a temporally frozen flow, the gains, relative to the na\"ive
strategy, justify this additional level of difficulty.

We use the adversarial $\mathcal{Q}$-learning approach outlined
above (parameter values in tab.~\ref{tab:Parameters}) to arrive at the optimal
scheme for path-planning in a 2D turbulent flow. To quantify the performance of
the smart microswimmers, we introduce equal numbers of smart (master-slave
pairs) and na\"ive microswimmers into the flow.  The scheme presented here pits
$\mathcal{Q}$-learning against the na\"ive strategy and enables the adversarial
algorithm to find a strategy that can out-perform the na\"ive one. (Without the
adversarial approach, the final strategy that is obtained may end up being
sub-optimal.)  


\begin{figure}[!ht]
	\includegraphics[scale=0.33]{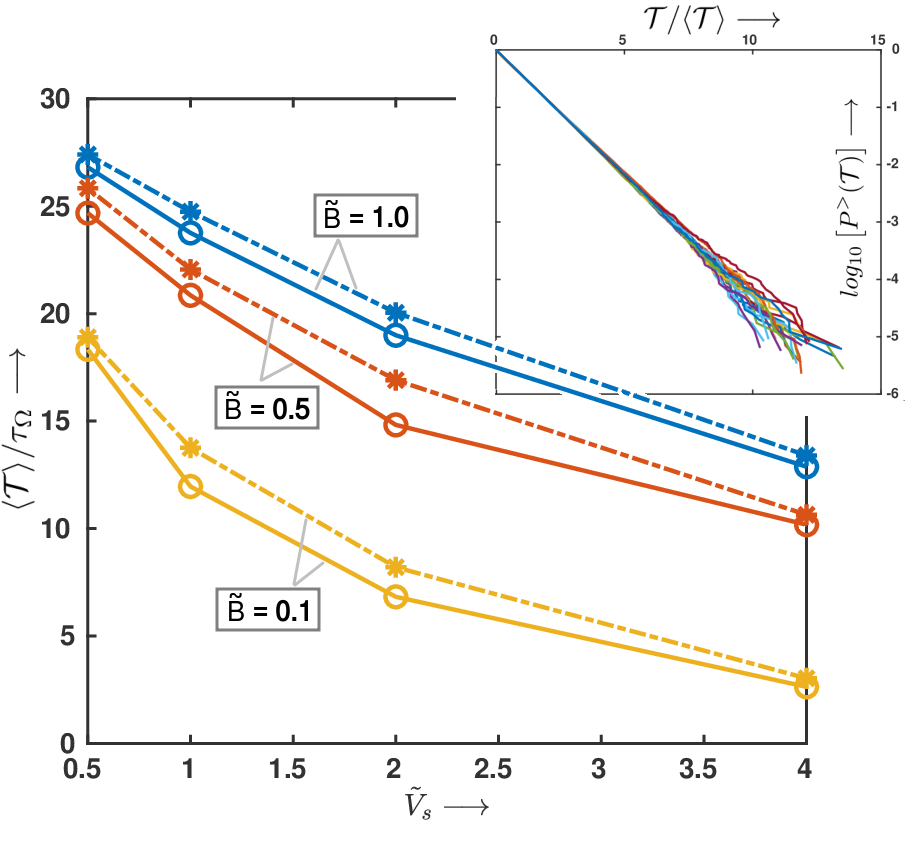}
	\caption{The dependence of $\langle T \rangle$ on $\tilde{V}_s$, for different values
	         of $\tilde{B}$, shown for the na\"ive strategy (dotted line) and 
		 for adversarial
	         $\mathcal{Q}$-learning (solid line), for our 2D turbulent flow. 
		 The plot shows that, in 
	         the parameter space that we have explored, our 
		 adversarial-$\mathcal{Q}$-learning 
	         method yields a 
	         lower value $\langle T \rangle$ than in the na\"ive strategy. 
		 The plot in the inset 
	         shows that the CPDF of $T$ has an exponential tail.}
	 \label{fig:T_VS_Vs}
\end{figure}

\section{Results}

\begin{figure}[!ht]
	\includegraphics[scale=0.18]{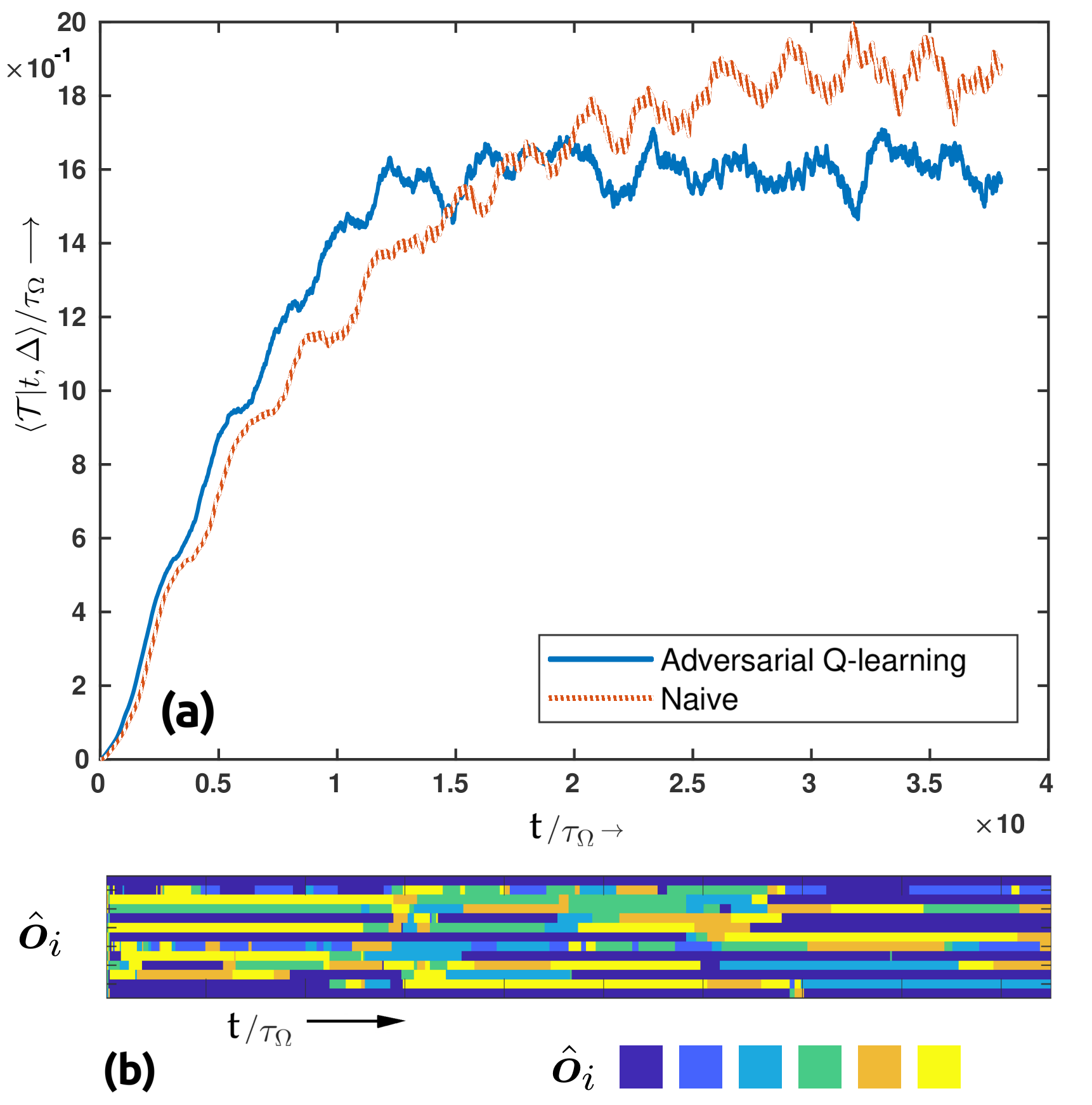}
	\caption{ Learning statistics in 3D: (a) The performance trend, $\langle
	          \mathcal{T} | t, \Delta\rangle / \tau_{\Omega}$, 
	         with $\Delta = 10 \tau_\Omega$ for adversarial $\mathcal{Q}$-learning
	         ({\color{blue} blue line}) and na\"ive strategy ({\color{red} red 
	         broken-line}) for microswimmers in a 3D homogeneous isotropic turbulent
	         flow, for $\tilde{V}_s = 1.5$ and $\tilde{B}=0.5$. 
	         The trend shows a slow rise in performance, similar to
	         that observed in 2D. In 3D the $\mathcal{Q}$-learning is performed by using
	         13 states and 6 actions defined in Appendix~\ref{sec:3D_State_Def};
	         (b) The evolution of $\hat{\mathbf{o}}_i$ in 3D shows that learning has not
	         reached a steady state due to lower probability of swimmers reaching the
	         target, compared to 2D case.}
	\label{fig:3D}
\end{figure}

The elements of $\mathcal{Q}$
evolve during the initial-training stage, so $P^>(\mathcal{T})$ also evolves in
time until the system reaches a statistically steady state (in which 
the elements of $\mathcal{Q}$ do not change). Hence, $\langle\mathcal{T}\rangle$ also
changes during the initial-training stage; to capture this time dependence, we define 
$\langle\mathcal{T}(t)\rangle := 1/N(t) \; \sum_{i=1}^{N(t)} \mathcal{T}_i$, where 
$\mathcal{T}_i$ is the time taken by the $i^{th}$ microswimmer, since its intialization, 
to arrive at its target at the time instant $t$ and $N(t)$ is the number of microswimmers 
that reach their targets at time instant $t$. We find that $\langle\mathcal{T}(t)\rangle$ 
shows large fluctuations; so we average it over a time window 
$\Delta$ and define $\langle \mathcal{T}|t,\Delta\rangle := 1/\Delta \;
\int_t^{t+\Delta} \langle \mathcal{T}(\tau)\rangle \; d\tau$.
The initial growth in $\langle \mathcal{T}|t,\Delta\rangle$ arises because
$\langle \mathcal{T}|t,\Delta\rangle \le t$.
The plots in figs.~\ref{fig:Performance}(a) and ~\ref{fig:3D} show the time
evolution of $\langle \mathcal{T}|t,\Delta\rangle$ for the smart and na\"ive microswimmers.
Note that $\hat{\mathbf{o}}_i$ becomes a constant, for large $t$, in 
fig.~\ref{fig:Performance}(b); this implies that the elements of $\mathcal{Q}$ have 
settled down to their steady-state values.

Figures~\ref{fig:Performance}(a), and~\ref{fig:Performance}(b) show the evolution of 
$\langle \mathcal{T} | t,
\Delta \rangle$ and $\hat{\mathbf{o}}$, respectively, for the na\"ive
strategy and our adversarial-$\mathcal{Q}$-learning scheme. After
the initial learning phase, the $\mathcal{Q}$-learning algorithm explores
different $\hat{\mathbf{o}}$, before it settles down to a steady state. It
is not obvious, \textit{a priori}, if there exists a stable, non-trivial,
optimal strategy, for microswimmers in turbulent flows, that could out-perform
the na\"ive strategy.  The plot in fig.~\ref{fig:T_VS_Vs} shows the
improved performance of our adversarial-$\mathcal{Q}$-learning scheme over the
na\"ive strategy, for different values of $\tilde{V}_s$ and $\tilde{B}$; in
these plots we use $\langle \mathcal{T} \rangle = \langle \mathcal{T} | t \rightarrow \infty,
\Delta \rangle$, so that the initial transient behavior in learning is
excluded. The inset in fig.~\ref{fig:T_VS_Vs} shows that $P^>(\mathcal{T})$
has an exponential tail, just like the na\"ive scheme in fig.~\ref{fig:Tail},
which implies the smart microswimmers also get trapped; but a lower value of
$\langle \mathcal{T} \rangle$ implies they are able to escape from the traps faster than
microswimmers that employ the na\"ive strategy. Note that the presence of a
possible noise in the measurement of the discrete vorticity $\mathcal{S}_\omega$ should 
not change our findings because of the coarse discretization we use in defining the states.

In a 3D turbulent flow, we also obtain such an improvement, with our
adversarial $\mathcal{Q}$-learning approach, over the na\"ive strategy. The details
about the 3D flows, parameters, and the definitions of states and actions are
given in Appendix~\ref{sec:3D_State_Def}. In fig.~\ref{fig:3D} we show a
representative plot, for the performance measure, which demonstrates this
improvement in the 3D case (cf.  fig.~\ref{fig:Performance} for a 2D
turbulent flow).

\section{Conclusions}

We have shown that the generic $\mathcal{Q}$-learning approach can be adopted
to solve control problems arising in complex dynamical systems.
In~\cite{HJB}, global information of the flows has been used for 
path-planning problems in autonomous-underwater-vehicles navigation to improve 
their efficiency, based on the Hamilton-Jacobi-Bellmann approach. In contrast, 
we present a scheme that uses only the local flow parameters for the path planning.

The flow parameters (tab.~\ref{tab:Flow_Parameters}) and the learning
parameters (tab.~\ref{tab:Parameters}) have a significant impact on the
performance of our adversarial-$\mathcal{Q}$-learning method. Even the choice
of observables that we use to define the states $(\mathcal{S}_{\omega},
\mathcal{S}_{\theta})$ can be changed and experimented with. Furthermore, the
discretization process can be eliminated by using deep-learning approaches,
which can handle continuous inputs and outputs~\cite{Deep_Q}. Our formulation
of the optimal-path-planning problem for microswimmers in a turbulent flow is a
natural starting point for detailed studies of control problems in turbulent
flows.

\section{Discussion}
We were made aware of~\cite{BB} during the writing of this manuscript, 
where they tackle the problem using an Actor-Critic reinforcement learning scheme.

We contrast, below, our reinforcement-
learning approach with that of Ref.~\cite{BB}.
\begin{itemize}

\item
Reference~\cite{BB} uses 900 discrete states, which are defined based on
the approximate location of the microswimmer. By contrast, our
scheme uses only the local vorticity ($\mathcal{S}_\omega$), at the position of the
microswimmer, and the orientation ($\mathcal{S}_\theta$); after discretization, we
retain only 12 states. In analogy with navigation parlance, Ref.~\cite{BB}
uses a GPS and our approach uses a light-house along with a
local-vorticity measurement.

\item
In Ref.~\cite{BB}, the states are sensed periodically and the elements of
$\mathcal{Q}$ are updated at every sensing instant. In constrast, we monitor
the states continuously and update the elements of $\mathcal{Q}$ only when
there is a state change. If the periodicity of sensing is smaller than
the rate of change in states of the microswimmer, both schemes
should show similar convergence behaviors.

\item
Reference~\cite{BB} uses a conventional, episode-based training approach,
which is sequential, whereas we use multiple microswimmers to
perform parallel training.
         
\item
Reference~\cite{BB} uses an actor-critic approach, whereas we use an
adversarial-learning method.         

\end{itemize}

\acknowledgments{
We thank DST, CSIR (India), BRNS, and the Indo-French Centre
for Applied Mathematics (IFCAM) for support.
}


\begin{thebibliography}{99}

\bibitem{Zermelo}
E. Zermelo {\"U}ber das navigationsproblem bei ruhender oder
  ver{\"a}nderlicher windverteilung,
\newblock {\em ZAMM-Journal of Applied Mathematics and Mechanics/Zeitschrift
  f{\"u}r Angewandte Mathematik und Mechanik} 11(2):114--124 (1931).

\bibitem{ML_FM}
S. Brunton, B. Noack, P. Koumoutsakos, {Machine Learning for Fluid Mechanics},
\newblock {\em arXiv e-prints} p. arXiv:1905.11075.

\bibitem{Soaring}
G. Reddy, A. Celani, T. J. Sejnowski, M. Vergassola, {Learning to soar in
  turbulent environments},
\newblock {\em Proceedings of the National Academy of Sciences}
  113(33):E4877--E4884 (2016).

\bibitem{PRL}
S. Colabrese, K. Gustavsson, A. Celani, L. Biferale, {Flow navigation by smart
  microswimmers via reinforcement learning},
\newblock {\em Physical review letters} 118(15):158004 (2017).

\bibitem{3D}
K. Gustavsson, L. Biferale, A. Celani, S. Colabrese, {Finding efficient
  swimming strategies in a three-dimensional chaotic flow by reinforcement
  learning},
\newblock {\em The European Physical Journal E} 40(12):110 (2017)


\bibitem{Dusenbery}
D. Dusenbery, {\em Living at Micro Scale: The Unexpected Physics of Being
  Small},
\newblock (Harvard University Press) (2009).

\bibitem{Durham}
W. M. Durham, et~al., {Turbulence drives microscale patches of motile
  phytoplankton},
\newblock {\em Nature communications} 4:2148 (2013).

\bibitem{Michalec}
F. G. Michalec, S. Souissi, M. Holzner, {Turbulence triggers vigorous swimming
  but hinders motion strategy in planktonic copepods},
\newblock {\em Journal of the Royal Society Interface} 12(106):20150158 (2015).

\bibitem{Fish}
S. Verma, G. Novati, P. Koumoutsakos, {Efficient collective swimming by
  harnessing vortices through deep reinforcement learning},
\newblock {\em Proceedings of the National Academy of Sciences of the United
  States of America} 115(23):5849—5854 (2018).

\bibitem{Barrows}
E. Barrows, {\em Animal Behavior Desk Reference: A Dictionary of Animal
  Behavior, Ecology, and Evolution, Third Edition},
\newblock (Taylor \& Francis) (2011).

\bibitem{RP_Review}
R. Pandit, et~al., {An overview of the statistical properties of
  two-dimensional turbulence in fluids with particles, conducting fluids,
  fluids with polymer additives, binary-fluid mixtures, and superfluids},
\newblock {\em Physics of Fluids} 29(11):111112 (2017).

\bibitem{Pedley}
T. J. Pedley, J. O. Kessler, {Hydrodynamic phenomena in suspensions of swimming
  microorganisms},
\newblock {\em Annual Review of Fluid Mechanics} 24(1):313--358 (1992).

\bibitem{Sutton}
R. S. Sutton, A. G. Barto, {\em Reinforcement learning: An introduction},
\newblock (Cambridge, MA: MIT Press) (2011).

\bibitem{Watkins}
C. J. Watkins, P. Dayan, {Technical note: Q-learning},
\newblock {\em Machine Learning} 8(3):279--292 (1992).

\bibitem{Survey}
L. P. Kaelbling, M. L. Littman, A. W. Moore, {Reinforcement learning: A survey},
\newblock {\em Journal of artificial intelligence research} 4:237--285 (1996).

\bibitem{Supp}
\newblock {\em See Supplementary material}.

\bibitem{canuto}
C. Canuto, M. Y. Hussaini, A. Quarteroni, T. A. Zang, {\em Spectral methods},
\newblock (Springer) (2006).

\bibitem{pramanareview}
R. Pandit, P. Perlekar, S. S. Ray, {Statistical properties of turbulence: an
  overview},
\newblock {\em Pramana} 73(1):157 (2009).

\bibitem{HJB}
D. Kularatne, S. Bhattacharya, M. A. Hsieh, {Optimal path planning in
  time-varying flows using adaptive discretization},
\newblock {\em IEEE Robotics and Automation Letters} 3(1):458--465 (2018).

\bibitem{Deep_Q}
T. P. Lillicrap, et~al., {Continuous control with deep reinforcement
  learning},
\newblock {\em arXiv preprint arXiv:1509.02971}.

\bibitem{BB}
L. Biferale, et~al., {Zermelo’s problem: Optimal point-to-point 
  navigation in 2D turbulent flows using reinforcement learning},
\newblock {\em Chaos} 29, 103138 (2019).

\bibitem{Watkins}
C. Watkins, {Learning from Delayed Rewards},
\newblock {\em PhD thesis, University of Cambridge, Cambridge, England} (1989).

\end{thebibliography}

\appendix
\clearpage
\section{Flowchart}
\label{sec:Flowchart}
\begin{figure}[!ht]
	\includegraphics[scale=0.19]{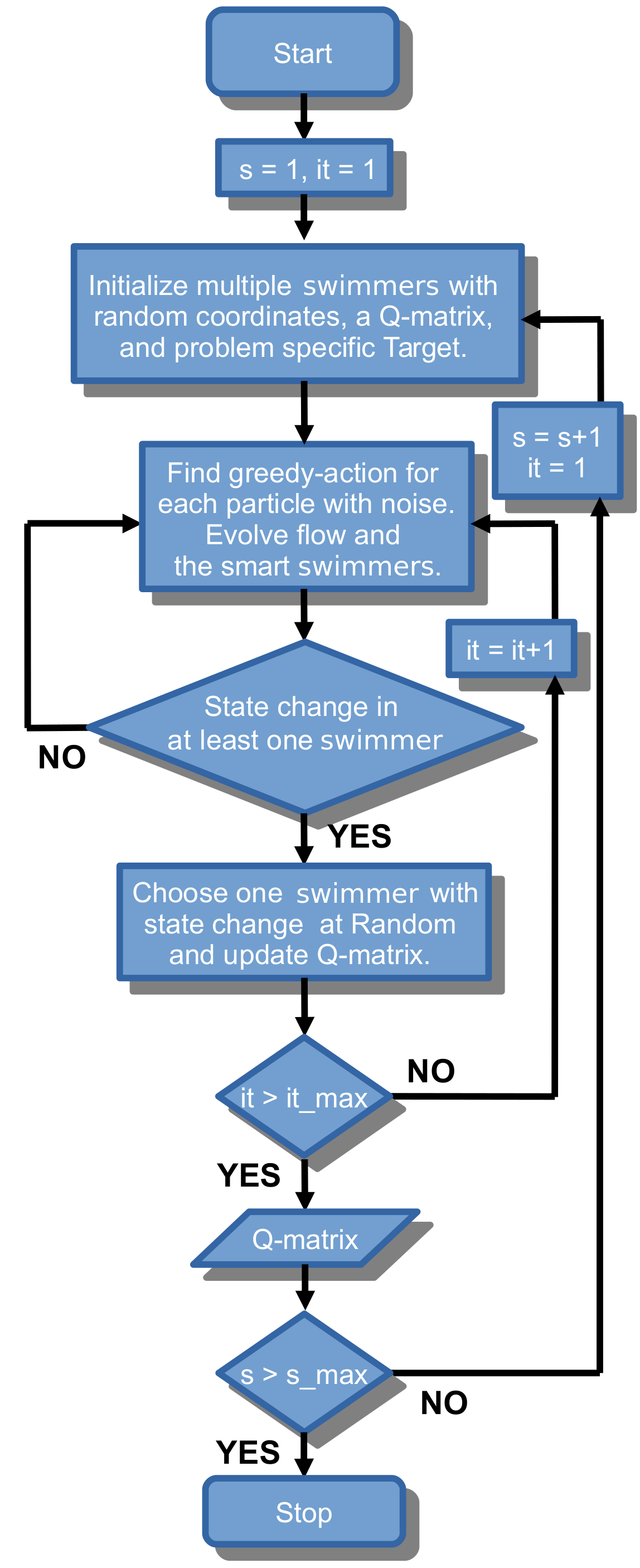}
	\caption{This flow chart shows the sequence of processes involved in our
	         adversarial $\mathcal{Q}$-learning algorithm. }
	\label{fig:Flowchart}
\end{figure}

Figure~\ref{fig:Flowchart} shows the sequence 
             of processes involved in our adversarial-$\mathcal{Q}$-learning scheme. Here 
             $it$ stands for the iteration number and $s$ is the number of sessions. We 
             use a greedy action in which the action corresponding to the maximum 
	         value in the $\mathcal{Q}$ matrix, for the state of the microswimmer, 
		 is performed; $\epsilon$-greedy step ensures with probability 
		 $\epsilon_g$ that the non-optimal action is chosen. Furthermore, we find that
	         episodic updating of the values on the $\mathcal{Q}$ matrix lead 
		 to a deterioration of performance; therefore, we use continuous 
		 updating of $\mathcal{Q}$.

\section{State and action definitions for 3D turbulent flow}
\label{sec:3D_State_Def}
\begin{figure}[!ht]
	\includegraphics[scale=0.2]{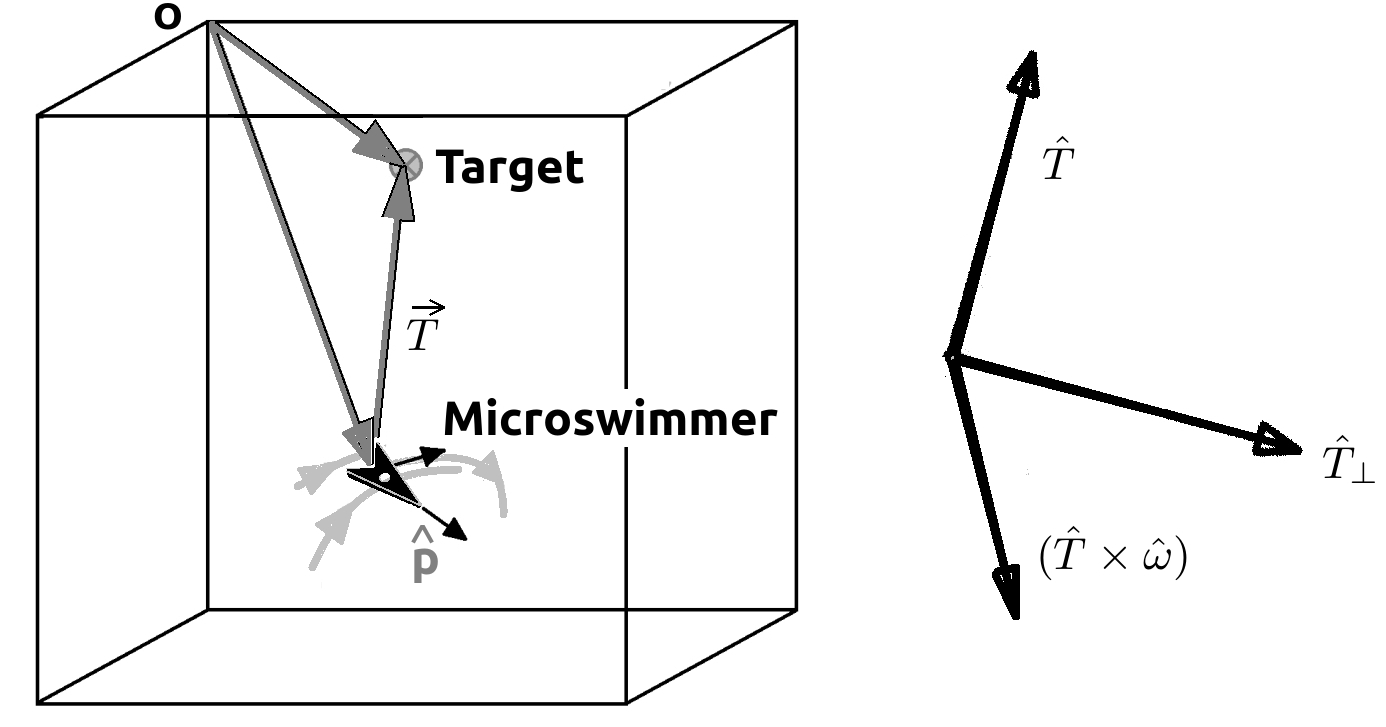}
	\caption{We define a Cartesian coordinate
	 system by using the ortho-normal triad $\left\lbrace \hat{\mathbf{T}}, (\hat{\mathbf{T}}\times
	\hat{\bm{\omega}}), \hat{\mathbf{T}}_\perp \right\rbrace$; thus, all the vectorial quantities
	are represented in terms of this observer-independent coordinate 
	system.}
	\label{fig:Sphere}
\end{figure}

\begin{figure*}[t]
	\includegraphics[scale=0.2]{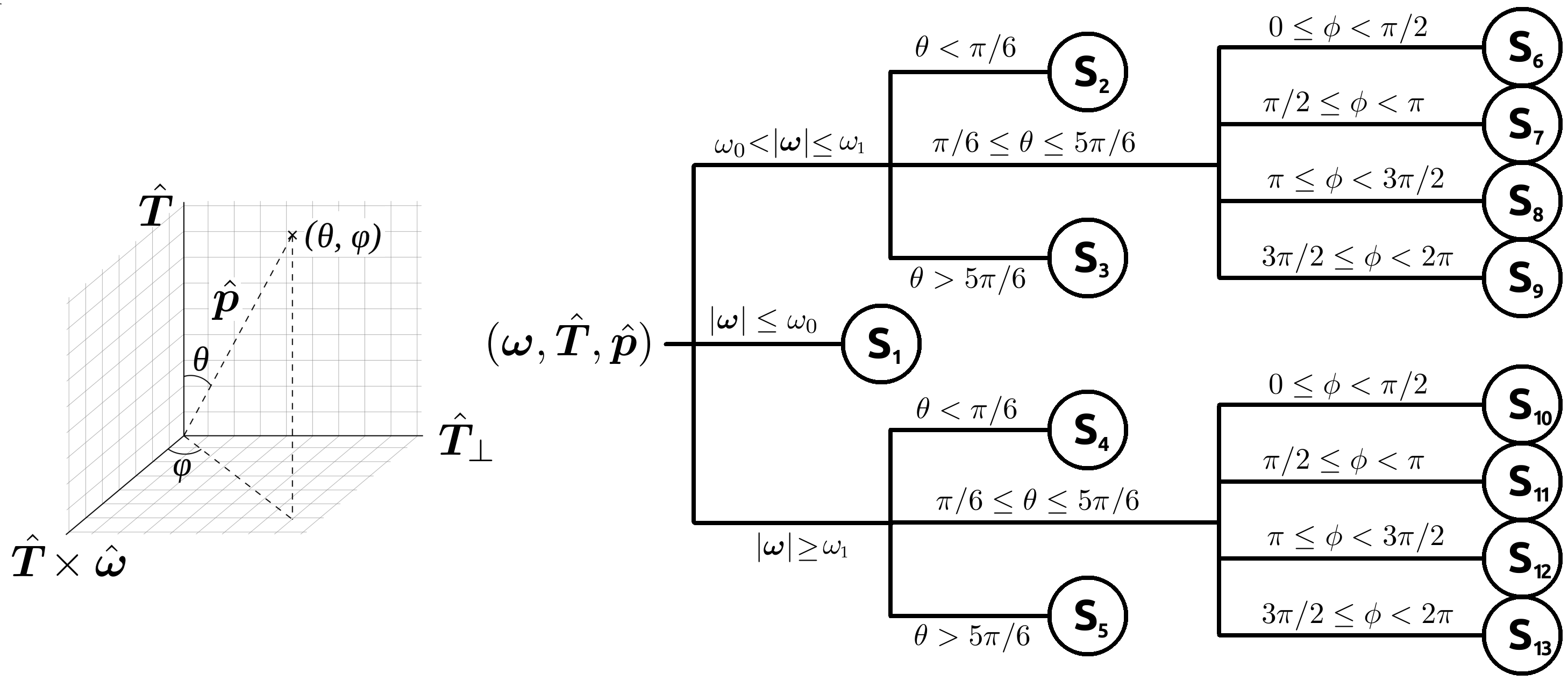}
	\caption{Discretization of states in 3D: We define a spherical-polar coordinate
	  system for each particle with the $z$ axis pointing along the $\hat{\mathbf{T}}$
	   direction and the
	   $x$ axis along $\hat{\mathbf{T}}_\perp$. We define the canonical angles $\theta$
	    and $\phi$, and discretize the states into 13, based on the
	    magnitude of $\vec{\bm{\omega}}$, where $\omega_0$ and $\omega_1$ are 
	    state-definition parameters (we use $\omega_0=\omega_{rms}/3$ and 
	    $\omega_1=\omega_{rms}$), 
	    and the direction of $\hat{\bm p}$, with respect to the triad, is defined in 
	    fig.~\ref{fig:Sphere}.}
	\label{fig:States}
\end{figure*}

From our DNS of the 3D Navier-Stokes equation we obtain a statistically steady, 
homogeneous-isotropic turbulent flow in a $128\times 128
\times 128$ periodic domain. We introduce passive microswimmers into
this flow. To define the states, we fix a coordinate triad, defined by 
$\left\lbrace \hat{\mathbf{T}}, (\hat{\mathbf{T}}\times \hat{\bm{\omega}}), \hat{\mathbf{T}}_{\perp} \right\rbrace$ 
as shown in fig. \ref{fig:Sphere}; here, $\hat{\mathbf{T}}$ is the unit vector pointing
from the microswimmer to the target, $\hat{\bm{\omega}}$ is the vorticity pseudo-vector, and 
$\hat{\mathbf{T}}_{\perp}$ is defined by the conditions $\hat{\mathbf{T}}_{\perp} \cdot \hat{\mathbf{T}} = 0$ and
$\hat{\mathbf{T}}_{\perp} \cdot (\hat{\mathbf{T}}\times\hat{\bm{\omega}}) = 0$.
This coordinate system is ill-defined if $\vec{T}$ is parallel to $\vec{\bm{\omega}}$.
To implement our $\mathcal{Q}$-learning in 3D, we define 13 states: $\mathcal{S} = 
(\mathcal{S}_{|\bm{\omega}|}, \mathcal{S}_\theta, \mathcal{S}_\phi)$ (see 
fig.~\ref{fig:States}); and 6 actions, $\mathcal{A} = \left\{ \hat{\mathbf{T}}, -\hat{\mathbf{T}}, 
(\hat{\mathbf{T}}\times \hat{\bm{\omega}}),  -(\hat{\mathbf{T}}\times \hat{\bm{\omega}}), \hat{\mathbf{T}}_{\perp}, 
-\hat{\mathbf{T}}_{\perp} \right\}$. Consequently, the $\mathcal{Q}$ matrix is an array of 
size $13\times6$.

\clearpage


\end{document}